\documentclass[twocolumn,aps,superscriptaddress,floats,showpacs]{revtex4}
\usepackage{epsfig}

\begin{document}
\title{Enhanced inverse bremsstrahlung heating rates in a strong laser field}
\author{A. Brantov}
\affiliation{Theoretical Physics Institute, Department of Physics, University of Alberta, Edmonton, Canada}
\author{W. Rozmus}
\affiliation{Theoretical Physics Institute, Department of Physics, University of Alberta, Edmonton, Canada}
\author{R. Sydora}
\affiliation{Theoretical Physics Institute, Department of Physics, University of Alberta, Edmonton, Canada}
\author{C. E. Capjack}
\affiliation{Department of Electrical Engineering, University of Alberta, Edmonton, Canada}
\author{V. Yu. Bychenkov}
\affiliation{P. N. Lebedev Physics Institute, Russian Academy of Science, Moscow, Russia}
\author{V. T. Tikhonchuk}
\affiliation{Centre des Lasers Intenses et Applications, Universit\'e Bordeaux 1, France}
\date{\today}

\begin{abstract}
Test particle studies of electron scattering on ions, in an oscillatory electromagnetic field have shown that standard theoretical assumptions of small angle collisions and phase independent orbits are incorrect for electron trajectories with drift velocities smaller than quiver velocity amplitude. This leads to significant enhancement of the electron energy gain and the inverse bremsstrahlung heating rate in strong laser fields. Nonlinear processes such as Coulomb focusing and correlated collisions of electrons being brought back to the same ion by the oscillatory field are responsible for large angle, head-on scattering processes. The statistical importance of these trajectories has been examined for mono-energetic beam-like, Maxwellian and highly anisotropic electron distribution functions. A new scaling of the inverse bremsstrahlung heating rate with drift velocity and laser intensity is discussed.
\end{abstract}
\maketitle

\section{Introduction}
The description of electron-ion collisions in the presence of a strong oscillatory electric field has been an essential part of every laser plasma interaction model. Processes such as collisional laser energy absorption, bremsstrahlung radiation emission and transport include electron scattering on ions in the presence of a high frequency electromagnetic field. Advances in laser technology, particularly those related to the generation of ultra-short laser pulses and progress in inertial confinement fusion studies have challenged our understanding of scattering processes over a wide range of conditions and plasma parameters. Nonetheless, heating rate calculations \cite{do,silin,bunkin,kw,kull,pert1,pert2,decker,shvets,mulser} in strong laser fields have involved simplifications that are equivalent to a Born approximation for electron trajectories in the Coulomb fields. This approximation, although well-accepted in the physics community, should be revisited in view of recently discovered nonlinear processes such as harmonic generation and above threshold ionization \cite{corkum0,brabec,ivanov} that have been explained as being due to strong modifications of electron orbits in the laser field during electron-ion scattering events. Similar nonlinear modifications of particle trajectories have recently been discussed in numerical studies of inverse bremsstrahlung heating rates \cite{fraiman,fraimannew}. Our paper continues this analysis; we examine classical electron trajectories that contribute to enhanced plasma heating \cite{fraiman} and discuss their importance for calculations of effective collision frequencies with certain classes of electron distribution functions. We also summarize existing theories of inverse bremsstrahlung heating and compare them with results of our test particle calculations. 

For the purpose of making comparisons, seminal works by Dawson and Oberman \cite{do} and Silin \cite{silin} and numerous subsequent publications on high field effects \cite{bunkin,kw,kull,pert1,pert2,decker,shvets,mulser} are classified into several broad categories: classical-mechanical analysis of binary electron-ion collisions in the dielectric approximation \cite{do,pert2,decker,shvets}; kinetic theory which uses the Landau-type collision integral in the oscillating electron field \cite{silin}; ballistic model (or the model of instantaneous electron-ion collision) \cite{bunkin,pert1,mulser}; "first principle" quantum-mechanical calculations \cite{bunkin,kw}; and, more recently, the quantum kinetic theory in the dielectric approximation \cite{kull}. All theories consider the two body scattering problem as a starting point and their predictions lead to remarkably similar expressions which may differ only by a logarithmic factor. 

The agreement between these approaches follows from two common characteristics inherent in all theories, i.e. the Born approximation is used for the electron orbits in a Coulomb field and the cross-section is assumed to have a weak dependence on the laser field phase. The first approximation, with notable exception of the low frequency approximation \cite{kw}, has been employed in most quantum mechanical calculations \cite{bunkin} where it applies to fast electrons with the velocity $v_0$ satisfying the condition $Z e^2 /\hbar v_0 \ll 1$ \cite{fedorov}, where $Ze$ is the ion charge. The Born approximation is equivalent to small angle scattering and small momentum exchange in classical kinetic theory studies \cite{do,silin,shvets}. This corresponds to a straight line electron trajectory approximation, that is also a part of ideal plasma collision theories that lead to the Landau or Balescu-Guernsey-Lenard operators \cite{ichimaru}. A simplified description of electron orbits removes the sensitivity of scattering processes to the initial phase of an electric field. However,it has been demonstrated \cite{fraiman} that  groups of electrons entering an ion interaction sphere at specific phase can significantly modify the collisional cross-section and the heating rates. 

Irregular electron trajectories in the combined, high frequency field of a laser and a Coulomb field of an ion have been found by Wiesenfeld \cite{num1}. His numerical studies have identified stochastic trajectories for electrons with quiver velocity larger than the initial drift velocity. This effect has been elaborated upon in detail in recent studies by Fraiman {\it et al.} \cite{fraiman,fraimannew}. In addition to irregular trajectories that involve quasi-capture of electrons in complicated orbits encircling ions, a phase space analysis has revealed the existence of initial conditions corresponding to large angle correlated collisions with anomalously large energy transfer from the field to particles. It was shown recently in Ref. \cite{jetp} that the effect of correlated collisions exists not only in the Coulomb potential but also in short-range potentials in the presence of a strong laser field.

There is a similarity between these processes and the interpretation of such atomic physics phenomena as multi-photon ionization and harmonic generation proposed by Corkum \cite{corkum0}. It was suggested that free electrons that are created in the process of tunnel ionization can be brought back to the original atom by the laser field and undergo multiple correlated collisions.  This process is sensitive to the phase of an electron in the oscillatory field and is responsible for processes such as high-harmonic generation, two-electron ejection, double ionization, etc. From the point of view of quantum mechanics, the high probability of secondary scattering was explained by the phenomenon of Coulomb focusing \cite{brabec,ivanov}. This is a process by which the Coulomb attraction of the ion compensates for the natural dispersion of the electron wave packet. A classical analog involves the slowly moving electron that is brought repeatedly into close proximity of an ion as it performs large amplitude oscillations in the laser field. Each encounter results in a small deflection of an electron trajectory toward the ion. These small changes accumulate into a deflection of an electron trajectory leading to a head-on collision producing large momentum change and causes an electron to leave the ion interaction sphere. An important question in the context of laser produced plasmas is determining under what circumstances such trajectories are statistically significant and can alter macroscopic heating rates. 

Our paper examines effects of correlated collisions and irregular scattering trajectories on the inverse Bremsstrahlung heating rate. By using mechanisms such as the Coulomb focusing \cite{corkum0,brabec}, the parachute effect \cite{fraiman} or quasi-capture \cite{num1}, we analyze strongly modified electron orbits and classify them in terms of their initial phase, impact parameters, and drift velocities. By means of test particle simulations we examine the statistical importance of these trajectories. They indeed strongly modify the heating rates for all three different energy distribution functions used in the simulations, i.e. mono-energetic, Maxwellian and anisotropic electron distribution functions.

As we show in this paper, there are large discrepancies between existing analytical theories and test particle simulations of heating rates. They warrant further investigations that may include molecular dynamics simulations \cite{dufty,batishchev}. Classical models of electron-ion interactions in molecular dynamics models include an effective potential \cite{uhlenbeck,deutsch} to prevent the collapse of such system due to electrostatic attraction and properly account for quantum diffraction effects at short distances. We will also perform test particle simulations with this effective potential and will compare them with Coulomb potential scattering studies.

The paper is organized as follows. In Sec. \ref{sec3} we summarize existing theories of inverse bremsstrahlung heating and present three representative expressions. Section \ref{sec2} presents a calculation model and gives examples of scattering orbits. Section \ref{sec4} describes the case of mono-energetic, beam-like electron distribution functions. We discuss different wave polarizations and field strengths. Heating rates for the case of a Maxwellian electron distribution are calculated in Section \ref{sec43}. Section \ref{sec42} deals with anisotropic electron distributions that are encountered in photo-ionized plasmas. Section \ref{sec5} contains a summary and conclusions.

\section{Description of the energy exchange in the laser field}\label{sec3}

In the introduction, we have identified several groups of analytical results
describing electron-ion collisions in the presence of a homogeneous oscillating electric field. This division is a convenient characterization for various formalisms that are used to calculate the heating rate for electrons. All theories lead to remarkably similar heating rates, particularly for energetic electrons and small angle collisions, where electron scattering is well described by the Born approximation. 

The Dawson-Oberman model \cite{do,pert2,decker,shvets} of the electron-ion correlation function  is based on the classical Born approximation of binary collisions and on an explicit average over random ion positions. The Coulomb field of the immobile ion of the charge $Ze$ is considered as a weak perturbation to the electron motion which is comprised of a drift velocity ${\bf v}_0$ along a straight trajectory and a quiver velocity ${\bf v}_E = e {\bf E}/m\omega$ in a laser field of amplitude ${\bf E}$ abd the frequency $\omega$. Here $-e$ and $m$ are the charge and the mass of the electron. The Dawson-Oberman  collision operator describes electron-ion scattering as an instantaneous event resulting in a small angle deflection of the electron trajectory and accounts for the screening effect of other electrons and the high frequency field in the dielectric approximation. The electron energy gain has been found in the weak field limit \cite{do}, and also for an arbitrary strength of the oscillating field \cite{decker} with a Maxwellian distribution of electrons \cite{do,decker} or with an arbitrary distribution function \cite{pert2,shvets}. Derivations by Pert \cite{pert2} and Shvets and Fisch \cite{shvets} arrive at similar expressions starting from the test particle, two body scattering problem. The energy gain for a given electron in a plasma with the ion density $n_i$ has the following form \cite{pert2,shvets}:
\begin{equation}\label{hrborn}
\frac{d \epsilon}{d t}\!=\!\frac{2n_i Z^2 e^4}{m v_0^2} \sum_{-l_{0}}^{+\infty} l \! \int \frac {d {\bf k}}{k^2} J_l^2 \! \left(\frac{{\bf k v}_E}{\omega} \right) {\bf v}_0 \frac {\partial}{\partial {\bf k}} \,\delta \left(\frac{{\bf k} \cdot {\bf v}_0}\omega -l \right)\,,
\end{equation}
where $J_l$ is the Bessel function and the integration over the transferred momentum ${\bf k}$ involves the upper cut-off limit at $k_{max}= m v_0^2/2Z e^2$, which corresponds to the assumption of small angle scattering. 

Silin \cite{silin} has calculated the high frequency nonlinear conductivity and the effective collision frequency for a fully ionized plasma by using a kinetic equation with a collision integral which accounts for small-angle scattering. In this approach, the high frequency electric field defines the cut-off limit of the impact parameter and the effective collision velocity, but does not affect the collision event itself. The Silin's collision frequency is in close agreement with the Dawson-Oberman model \cite{decker} in both limits of weak and strong amplitude high frequency fields. A similar result for the electron energy gain has been obtained from the solution to the electron kinetic equation with the Landau collision operator after averaging the result with respect to the high frequency oscillations by Catto \cite{catto}.

Several classical models of the effective collision frequency have been derived from explicit treatment of the dynamics of electron-ion scattering assuming instantaneous and elastic interactions \cite{bunkin,pert1,mulser}. We will identify them as the impact approximation \cite{pert1,bunkin}, or the ballistic model \cite{mulser}. From the conservation of energy for such collisions, one finds that the change in the average electron kinetic energy, which is related to the electron drift velocity, is proportional to the change in the electron momentum. The energy gain is averaged with respect to the phase of laser field at the time of the electron-ion collision and expressed in terms of the transport cross section:
\begin{equation}\label{hrimpact}
\frac{d \epsilon}{d t} = \frac{2 n_i Z^2 e^4 }{m} \!\! \int_0^{2 \pi}\!\!\!\!\! d \phi \frac{\Lambda ({\bf v}_E \cdot {\bf v}_0 \,   \cos \phi + v_E^2 \cos^2 \phi)}
{(v_0^2 + 2 {\bf v}_E \cdot {\bf v}_0 \,   \cos \phi + v_E^2 \cos^2 \phi)^{3/2}}  \,.
\end{equation}
Here, the  Coulomb logarithm 
\begin{equation}\label{eqLambda}
\Lambda = \ln \left (1+\rho_{max}^2/\rho_{min}^2 \right)^{1/2}\,
\end{equation}
depends on the full electron velocity ${\bf v} =  {\bf v}_0 + {\bf v}_E \cos \phi$ \cite{mulser,pert3}  and $\rho_{max}=v/ \omega$ and $\rho_{min}=\max \{2 Ze^2/m v^2,\hbar /mv \}$ are the classical cut-offs at large and small impact parameters. According to Eq. (\ref{hrimpact}), there is a singularity in the integrand if the quiver velocity, $v_E$, approaches the drift velocity, $v_0$. This singularity corresponds to orbits passing close to the ion and first appears for electrons launched in the direction of the electric field. Correlated collisions in presence of high frequency electric field have been observed in numerical simulations of particle trajectories \cite{fraiman,num1} for the parallel launch, ${\bf v}_0 \parallel {\bf v}_E$. They remove this singularity and lead to an electron energy gain that is much larger than one that follows from Eq. (\ref{hrimpact}), if $v_0 < v_E$.

A quantum-mechanical description of electron-ion collisions in the Coulomb and laser fields employing the low frequency approximation has been developed by Bunkin and Fedorov \cite{bunkin} and Kroll and Watson \cite{kw}. Agreement between the quantum mechanical and classical derivations of the electron heating rate has been demonstrated by Ferrante {\it et al.} \cite{ferrante}. Within the theoretical framework of the Born approximation, this is a consequence of the well-known fact \cite{landau3} that the electron scattering cross section in the Coulomb field is the same in classical and quantum-mechanical calculations. The energy gain averaged with respect to the Maxwellian electron velocity distribution function \cite{schlessinger} also shows a good agreement between classical and quantum results. The recent quantum theory of the inverse Bremsstrahlung heating rate \cite{kull}, based on the dielectric approximation, compares well with classical results, provided the quantum cut-off parameters are introduced into the classical kinetic theory. 

Quantum mechanical calculations \cite{bunkin,schlessinger} describe the energy exchange rate between an electron and a laser field by using the cross sections derived by Kroll and Watson \cite{kw} for electron-ion scattering. The result is given by
\begin{eqnarray} \label{hrkw}
\frac {d \epsilon}{d t} \!&=&\! \frac{n_i Z^2 e^4}{2m v_0}\!\!\! \sum_{l=-l_{min}}^{+\infty}\!\!\!\!\! \int d {\bf n}^\prime \frac{ \xi l \,(1+\xi l)^{1/2} }{[1+\xi l/2-(1+\xi l)^{1/2} {\bf n \cdot n}^\prime ]^2} \times \nonumber \\
&& J_l^2 \left(\frac{2{\bf v}_E}{\xi v_0}[(1+\xi l)^{1/2}{\bf n}^\prime-{\bf n }] \right)\,, 
\end{eqnarray}
where ${\bf n}$ and ${\bf  n}^\prime$ refer to the propagation directions of the electron before and after collision, respectively, $\xi=2\hbar \omega / m v_0^2$, $v_0$ is the initial drift velocity, and $l_{min}=1/\xi$ is the maximum number of emitted photons. Energy exchange in this equation is expressed in terms of the differences between the total absorption cross section (the terms with $l>0$ in the sum) and emission (the terms with $l<0$) of $l$ photons.

\section{Computational model}\label{sec2}
Our simulations of non-relativistic electron scattering in the field of an ion and in a uniform laser field  are similar to studies of other authors \cite{fraiman,num1}. We solve Newton's equation for the motion of a test electron
\begin{equation} \label{eqN}
m \ddot{\bf r} = -e {\bf E} \sin \omega t - {\bf \nabla} U(r),
\end{equation}
where $U(r)$ is the electrostatic potential describing the electron-ion interaction calculated along the electron trajectory, ${\bf r}(t)$. The electron velocity ${\bf u}=\dot{\bf r}$ is divided into two components: the quiver velocity, with amplitude ${\bf v}_E$, and the drift velocity ${\bf v}$ and therefore ${\bf u} = {\bf v}- {\bf v}_E \cos\omega t$. In the absence of an oscillatory field and for the Coulomb potential, $U(r)=Ze^2/r$, the electron follows the Kepler orbit that is uniquely defined by its initial drift velocity ${\bf v}_0$ and the impact parameter $\rho$. The scattering process for the momentum exchange is described by the Rutherford cross-section, which corresponds to the surface area of the interaction radius $\rho_0 = Ze^2/mv_0^2$ multiplied by a logarithmic factor which accounts for small angle collisions. The limits in the logarithm are defined by the Debye screening length $\lambda_D=\sqrt{T_e/4\pi e^2n_e}$ at large distances (here $T_e$ is the electron temperature) and $\rho_0$. The small impact parameters correspond to large angle scattering events, including head-on electron-ion collisions, which contribute to the Rutherford cross section with finite probabilities. Similar properties describe electron-ion scattering in a weak oscillatory field $v_E/v_0 \ll 1$, the only exception being that the long distance cut-off is defined by the length $v_0/\omega$ if it is smaller than the Debye length.

The classical treatment of collisional absorption and particle heating gives a result, which within logarithmic accuracy, agrees with quantum mechanical calculations in the Born approximation \cite{schlessinger,kull}. The exact agreement between the quantum scattering problem and the Kepler orbit solutions is well known and is a unique feature of the Coulomb potential \cite{landau3}. In numerical studies based on classical mechanics such as molecular dynamics \cite{dufty} or  test particle simulations, the quantum diffraction effect at short distances can be introduced by imposing a cut-off in the Coulomb potential \cite{deutsch} at the electron de Broglie length, $\lambda_B=2\pi \hbar/mv$. This cut-off was first suggested by Uhlenbeck \cite{uhlenbeck}. It has allowed the successful application of classical theories to weakly degenerate systems \cite{ichimaru1}. For the simplest problem of electron scattering off a bare ion, no short range cut-off is necessary in view of the agreement between quantum and classical calculations. No such general agreement exists, however, when the oscillatory field modifies the scattering orbits. For example, the quantum theory of inverse bremsstrahlung in the Born approximation \cite{kull} produces limiting expressions with logarithmic terms involving a cut-off at $\lambda_B$. Also, there is no quantum theory going beyond the Born approximation \cite{comment}. Thus the problem of quantum corrections to the effective interaction potential is a difficult one and has no clear solution so far.  

\begin{figure}[!ht]
\epsfig{figure=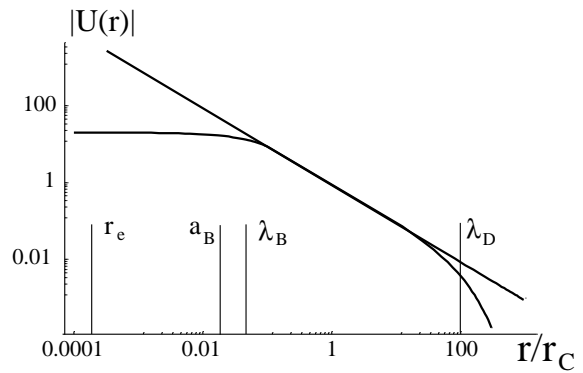, width=7.5cm}
\caption{Radial dependence of the effective potential $U(r)$ (\protect \ref{poten}) normalized by $Z e^2/r_C$ in comparison with the bare Coulomb potential, $r_C/r$. Examples of characteristic distances are shown in relation to $r_C=\sqrt{Ze/E}$ for $Z=10$: the Debye screening length, $\lambda_D$, was calculated for the electron temperature 1 keV and density $10^{20}$ cm$^{-3}$; the electron de Broglie length,  $\lambda_B$, was calculated for the laser intensity $10^{16}$ W/cm$^2$ and the wavelength 0.25 $\mu$m; the Bohr radius, $a_B$, and the classical electron radius, $r_e$.}\label{fig1}
\end{figure}

For these reasons, the potential used in our simulations accounts for cut-offs at both short and large distances: 
\begin{equation} \label{poten} 
U(r) = \frac{Z e^2}r \left [ 1- \exp \left (- \frac r{\lambda_B} \right ) \right] \exp \left (- \frac r{\lambda_D} \right)\,.
\end{equation}
The two lengths, $\lambda_B$ and $\lambda_D$, are considered as free parameters that allow one to make contact with known limiting cases. Figure \ref{fig1} compares the effective potential (\ref{poten}) with the Coulomb potential. The characteristic spatial scale of the problem is the distance $r_C=\sqrt{Ze/E}$ \cite{fraiman}, where the strength of the Coulomb field is the same of the oscillating field.  Here, the Debye length was calculated for an electron temperature of 1 keV and an electron density $10^{20}$ cm$^{-3}$. The de Broglie length, $\lambda_B=2\pi \hbar/mv_E$, was calculated at the quiver velocity $v_E$ corresponding to a laser intensity of $10^{16}$ W/cm$^2$ and a wavelength of 0.25 $\mu$m.  In our study, we examine, within the classical formulation, scattering orbits that clearly display non-perturbative modifications to the Kepler trajectories.  We proceed with two dynamical models involving the Coulomb potential, as in \cite{fraiman}, and the effective potential (\ref{poten}) with $\lambda_B$ calculated by using the quiver electron velocity. The upper cut-off at $\lambda_D$ is not important, because the correlated collisions occur for relatively small impact parameters and do not demonstrate the logarithmic divergence at large distances which is the characteristic feature of scattering in the Coulomb potential.

\begin{figure}[!ht]
\epsfig{figure=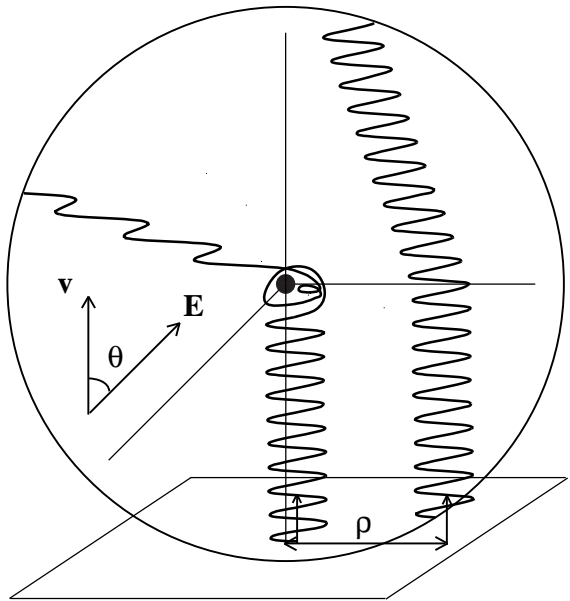, width=7.5cm}
\caption{Schematic electron orbits in the problem of electron-ion scattering in the laser field. Two typical electron trajectories are shown which originate at the launching plane. The sphere represents the long range cut-off of the effective potential.} \label{fig2}
\end{figure}

Numerical solutions of the equations of motion, Eq. (\ref{eqN}), require special attention regarding the choice of a suitable time step in computations of electron orbits near the scattering center. We have used the symplectic integration algorithm \cite{candy}, which ensures a good numerical accuracy and an exact conservation of dynamical invariants. Even though the test particle problem described by Eq. (\ref{eqN}) appears simple, the construction of the statistical ensemble of results for a broad range of initial conditions for the average energy gain poses considerable challenges in terms of computer memory and time requirements. The overall geometry of our simulations is shown in Fig. \ref{fig2}. Electrons were launched from the plane of incidence with a prescribed impact parameter $\rho$ and an initial drift velocity $v_0$ which was always normal to the plane. The interaction sphere was defined by the screening length, $\lambda_D$. The electron trajectories depend on the angle of ${\bf v}_0$ with respect to the high frequency field polarization and on the field phase. 

The relevance of individual scattering trajectories in calculations of the average energy gain follows from the validity of the two-body collision approximation. This is also the basic assumption of all theoretical results used here for comparison. The two-body scattering model requires that the given electron can only interact with one particular ion during the scattering event. Therefore, the amplitude of electron oscillations, $r_E=eE/m \omega^2$, in the laser field must be smaller than the average distance between ions $d=(3/4 \pi n_i)^{1/3}$.  This condition, $r_E<d$, reads
\begin{equation} \label{vcon1}
5.7 \times 10^{-14} \lambda_0^2 \sqrt{I_0} n_i^{1/3} < 1,
\end{equation}
where the laser wavelength $\lambda_0$ is in $\mu$m, the intensity $I_0$ is in W/cm$^2$ and the ion density is in cm$^{-3}$. For example, a laser pulse of intensity $I_0 \alt 10^{17}$ W/cm$^2$ and wavelength $\lambda_0 = 0.25 \,\mu$m satisfies the condition (\ref{vcon1}) in a gaseous plasma of $n_i = 10^{18}$ cm$^{-3}$.

\begin{figure}[!ht]
\epsfig{figure=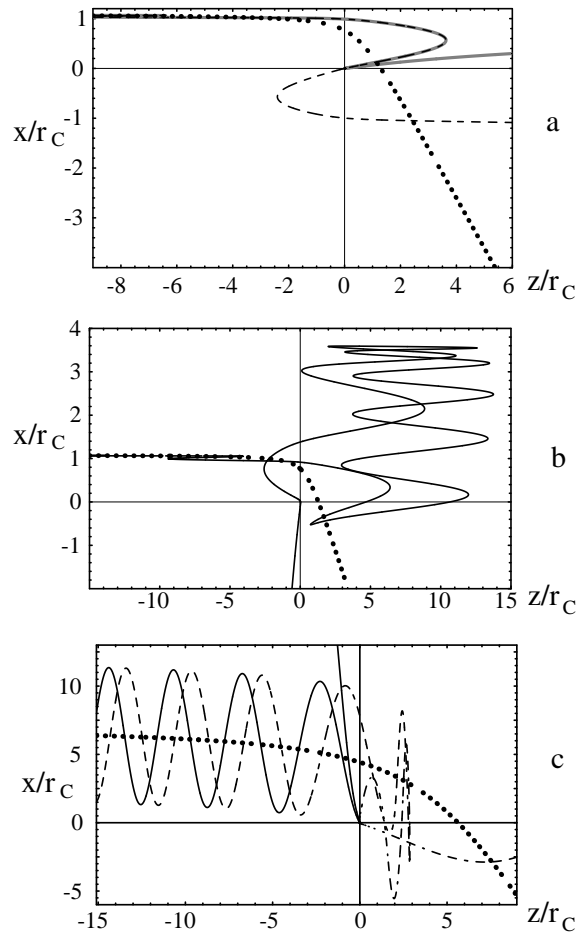, width=7.5cm}
\caption{Examples of electron trajectories: (a) A hyperbolic orbit in the absence of the oscillatory field (dotted line) is compared with orbits for particles with identical initial conditions and parallel velocities with respect to the field polarization for the Coulomb potential (solid gray line) and for the effective potential (dashed line). (b) A complicated electron orbit showing the quasi-trapping event -- a long time trapping around the ion (solid line) for the parallel launch geometry, with the impact parameter equal to $2.5 \rho_0$ and the initial drift velocity $v_0= 0.3 v_E$. This is compared with the hyperbolic orbit in the absence of the oscillatory field (dotted line). (c) Comparison of two similar trajectories of slightly different initial phases for the perpendicular launch geometry. In this case the impact parameter is $15 \rho_0$ and the initial drift velocity $v_0= 0.1 v_E$. The relevant Kepler orbit is shown as a dotted line. The remaining parameters for all examples are: $Z=10$, the laser wavelength 0.25 $\mu$m and the intensity $4.43 \cdot 10^{16}$ W/cm$^2$.} \label{fig3}
\end{figure}

The scattering trajectories that give rise to an enhancement of the electron energy gain \cite{fraiman} are illustrated in Fig. \ref{fig3}. Figure \ref{fig3}a compares three cases with the same initial conditions: a hyperbolic orbit in the absence of an oscillatory field (dotted line), a trajectory in a Coulomb field and in an oscillatory field (dashed line) and a solution to the equation of motion with an effective potential (\ref{poten}) and an oscillatory field (continuous line). In this example, the oscillatory field changes the trajectory and causes a head-on collision with the ion. This dramatic deflection is moderated by the quantum diffraction effect accounted for in the effective potential (\ref{poten}). The single trajectory of Fig. \ref{fig3}b corresponds to the case where the initial drift velocity is parallel to the direction of the oscillatory field. It illustrates the case of quasi-capture for the motion in the effective potential. The time dependent field reduces the instantaneous electron energy to a negative value allowing for a complicate bounded motion in the vicinity of an ion. The third example, Fig. \ref{fig3}c, displays two trajectories of electrons launched in the transverse direction to the oscillatory field with a small difference in the initial phase. At first, both orbits are very close to each other, yet after the collision they diverge dramatically. This sensitivity to phase and a strong modification of the scattering trajectory after an almost head-on collision, are manifestations of the effect of Coulomb focusing described in Refs. \cite{corkum0,brabec,ivanov}.

Our calculations focus on the energy gain of a single electron $\langle\epsilon\rangle$ for a given impact parameter averaged over the field phase and the rate of energy gain $d\epsilon/d t$ for a single electron averaged over the impact parameters and the field phase. These quantities were calculated directly from the simulation results according to the following expressions:
\begin{equation}\label{eq2}
\langle\epsilon\rangle = \frac m2 \, \langle v_0^{\prime 2}-v_0^2\rangle\,, \qquad \frac{d\epsilon}{d t}= n_i v_0 \int \langle\epsilon\rangle \,d^2 \rho\,,
\end{equation}
where ${\bf v}_0$ and ${\bf v}_0^\prime$ are the electron drift velocities before and after the collision (cf. Fig. \ref{fig2}) and the angular brackets denote the average over the field phase. The electron energy gain from the oscillatory electric field (\ref{eq2}) is compared with known analytical and numerical results.

In the following three sections we describe the results of our test particle simulations of the electron energy gain for different initial electron distribution functions. We start with the beam-like mono-energetic electron distribution function, followed by an isotropic Maxwellian distribution and finally with an anisotropic electron distribution function. 

\section{Energy gain for a mono-energetic electron beam}\label{sec4}
Here, we analyse the energy exchange between the oscillatory field and electrons of a given initial velocity. The ensemble of different trajectories are constructed by varying the initial phase and the impact parameter. In particular, the parallel and perpendicular polarization of the field with respect to the initial direction of the drift velocity are considered.

\subsection{Analytical expressions}\label{sec4a}
The geometry of a parallel field polarization allows further simplification of the expressions for the electron energy gain obtained by Dawson and Oberman, Eq. (\ref{hrborn}):
\begin{eqnarray}\label{bornpar}
&&  \frac {d \epsilon} {d t}= \frac {4 \pi n_i Z^2 e^4}{m v_0} \sum_{l=1}^{\infty} \left\{ \frac {2 J_l^2(l v_E/v_0)}{1+ l^2 \hbar^2 \omega^2/m^2 v_0^4} - \right . \nonumber \\ && \left . \frac{lv_E}{v_0} J_l \left(\frac{l v_E}{v_0}\right) \left[J_{l-1}\left(\frac{l v_E}{v_0 }\right)-J_{l+1} \left(\frac{ lv_E}{v_0} \right )\right] \times \right .
 \nonumber \\
&&  \left. \ln \left ( 1+\frac {m^2 v_0^4}{l^2 \hbar^2 \omega^2} \right )
\right \}\,, 
\end{eqnarray}
and Kroll and Watson, Eq. (\ref{hrkw}): 
\begin{equation} \label{kwpar}
\frac {d \epsilon} {d t} =\frac {\pi n_i Z^2 e^4}{m v_0}\!\!\! \sum_{l=-l_{min}}^{+\infty} \! \int \limits_{1-\sqrt{1+ \xi l}}^{1+ \sqrt{1+ \xi l}} \! \frac{d y \, \xi l }{(y+\xi l/2)^2} J_l^2\left(\frac{2v_E y}{v_0 \xi}\right) \,.
\end{equation}
Equation (\ref{bornpar}) follows from Eq. (\ref{hrborn}) after integration over the transfered momentum ${\bf k}$ with the upper limit cut-off at $k_{max}= m v_0/\hbar$, which corresponds to quantum short distance cut-off at the de Broglie length calculated at the initial electron drift velocity. The parameter $\rho_{min}= \hbar/m v_0$. A more detailed study of the quantum mechanical expression (\ref{kwpar}) can be found in Ref. \cite{daniel}, where it is shown that an enhancement of the total cross-section occurs for $v_0 \sim v_E$ and an oscillatory behavior is observed for $v_0 < v_E$. Note, that in the classical limit, $\xi l \ll 1$, Eq. (\ref{kwpar}) gives an expression which is similar to Eq. (\ref{bornpar}).

In the limit of a weak electric field, where  $v_E/v_0 \ll \xi \ll 1,$ one recovers from Eq. (\ref{kwpar}) the well known expression
\begin{equation}\label{eq11}
\frac {d \epsilon} {d t} = \frac{4 \pi n_i Z^2 e^4 v_E^2}{m v_0^3} \ln \left(\frac{2 m v_0^2}{\hbar \omega}\right)\,
\end{equation}
which accounts for the quantum cut-off at small distances and the high frequency cut-off, $\rho_{max}= v_0/\omega$, at large distances.

There is no easy way of simplifying theoretical results for the perpendicular polarization of the field. We use the original expressions (\ref{hrborn}) and  (\ref{hrkw}) for the comparisons with our numerical results.

\subsection{Simulation results}\label{sec4s}
\begin{figure}[!ht]
\epsfig{figure=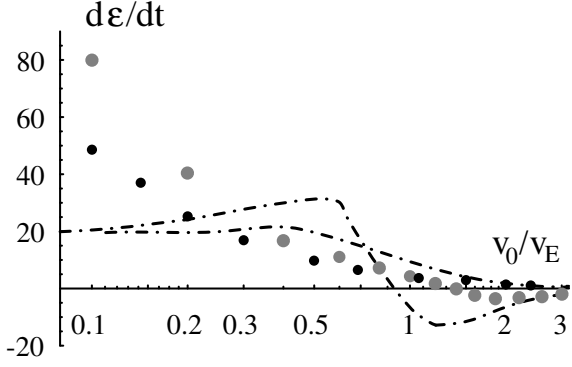, width=7.5cm}
\caption{The energy gain rate $d \epsilon / d t$ normalized to $n_i Z^2 e^4/m v_E$ as a function of the normalized initial drift velocity for weak field values, $r_E \alt \rho_E$. The parameters are: $Z=10$, the laser wavelength 0.25 $\mu$m and the intensity $1.8 \times 10^{15}$ W/cm$^2$. Dots describe numerical results for the parallel launch case (gray dots) and the perpendicular launch case (solid dots). They are compared with the Kroll-Watson theory (dash-dot curves). Theoretical results showing negative energy gains correspond to the parallel launch.} \label{fig4}
 \end{figure}

\begin{figure}[!ht]
\epsfig{figure=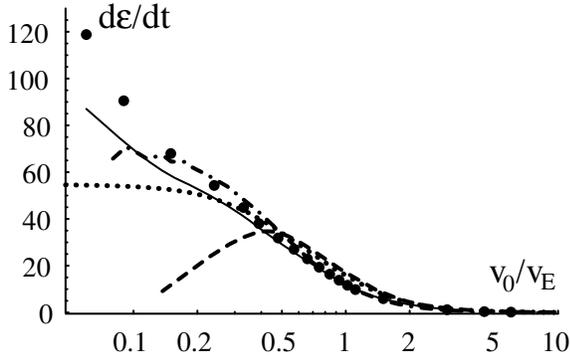, width=7.5cm}
\caption{The energy gain rate $d \epsilon / d t$ normalized to $n_i Z^2 e^4/m v_E$ as a function of the normalized initial drift velocity for the perpendicular launch geometry. The parameters of the simulations are: $Z=10$, the laser wavelength 0.25 $\mu$m and the oscillatory field intensity is equal to $4.43 \times 10^{16}$ W/cm$^2$.  Numerical results for the Coulomb potential shown as large dots and for the effective potential as the solid curve; the Kroll-Watson approximation -- dash-dotted line, the Dawson-Oberman theory -- dashed line, the classical approach -- dotted line.}\label{fig5}
\end{figure}

\begin{figure}[!ht]
\epsfig{figure=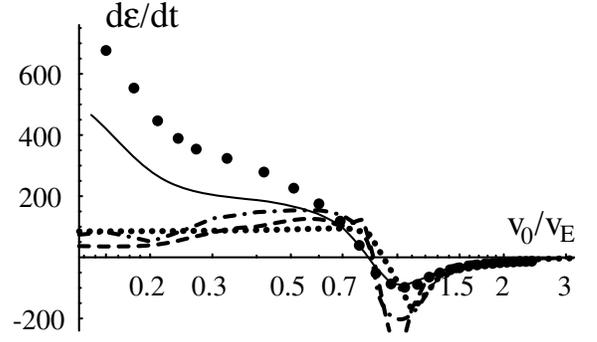, width=7.5cm}
\caption{The energy gain rate $d \epsilon / d t$ normalized to $n_i Z^2 e^4/m v_E$ for the parallel launch geometry. The parameters are: $Z=10$, the laser wavelength 0.25 $\mu$m and the intensity is equal to $4.43 \times 10^{16}$ W/cm$^2$. Numerical results for the Coulomb potential are shown as large dots and for the effective potential as the solid curve; the Kroll-Watson approximation -- dash-dotted line, the Dawson-Oberman theory -- dashed line, the classical approach -- dotted line.} \label{fig6}
 \end{figure}

The results of numerical simulations are illustrated by Figs. \ref{fig4}, \ref{fig5}, and \ref{fig6}. They display the average electron energy gain, $d \epsilon/dt$ (\ref{eq2}), as a function of initial drift velocity normalized to $v_E$. Different $v_0/v_E$ values along horizontal axis were achieved by changing the initial drift velocity and keeping $v_E$ constant. In addition to $v_0/v_E$, the results also depend on the strength of the electric field. We characterize this dependence by using the parameter $r_E/\rho_E$, where $\rho_E=Ze^2/mv_E^2=\rho_0 v_0^2/v_E^2$ is the distance for a large angle scattering event for an electron with the velocity $v_E$. It replaces $\rho_0$ in the definition of scattering cross-section for trajectories with $v_0/v_E <1$. The parameter $r_E/\rho_E$ can be conveniently presented as
\begin{equation}\label{rre}
    r_E/\rho_E= 3.4 \times 10^{-20} I_0^{3/2} \lambda_0^4/Z
\end{equation}
where the laser intensity is in W/cm$^2$ and the wavelength is in $\mu$m. It is large for typical conditions of present laser-plasma interaction experiments. However, the limit of small values of this parameter is also instructive to consider.

Figure \ref{fig4} corresponds to $r_E/\rho_E \alt 1$ while Figs. \ref{fig5}, \ref{fig6} describe the high intensity limit, $r_E \gg \rho_E$. As can be seen from Figs. \ref{fig3}, that are also drawn for $r_E/\rho_E \gg 1$, scattering events of electrons experiencing large amplitude oscillations in the electric field are very sensitive to phase and are likely to result in large energy gains but only for a selected number of particles. Figure \ref{fig5} shows numerical results for the energy gain (\ref{eq2}) with the Coulomb potential (large dots) and the effective potential $U(r)$ (\ref{poten}) (continuous line) for the transverse launch (${\bf v}_0 \perp {\bf v}_E$) case. They are compared with results of the Dawson-Oberman theory (\ref{hrborn}) (dashed line), the classical model (\ref{hrimpact}) (dotted line), and the Kroll-Watson theory (\ref{hrkw}) (dash-dotted line). The energy gain (\ref{eq2}) for the parallel launch (${\bf v}_0 \| \,{\bf v}_E$) is compared with results of three theoretical models, (\ref{hrimpact}), (\ref{bornpar}), and (\ref{kwpar}) in Fig. \ref{fig6}.

Two main features characterize the numerical results in Figs. \ref{fig4}, \ref{fig5}, and \ref{fig6}: an agreement between theoretical and numerical results for high velocities, $v_0 \gg v_E$, and increasing values of the energy gain for small initial velocities,  $v_0 \ll v_E$ which is in strong disagreement with all theoretical predictions. An agreement at large drift velocities follows from the applicability of the Born approximation to these trajectories. All curves in Figs. \ref{fig4}, \ref{fig5}, and \ref{fig6} are well approximated by a classical result predicting the decrease of the energy gain as $v_0^{-3}$. At small drift velocities, the calculated energy gain becomes larger. The increase is more dramatic for large fields, $r_E/\rho_E \gg 1$, and for trajectories calculated with the Coulomb potential (large dots) as compared to the effective potential (\ref{poten}) (solid lines) in Figs. \ref{fig5} and \ref{fig6}. 

\subsection{Weak field limit}\label{sec4w}
We identify a weak field case by $r_E/\rho_E \alt 1$. According to Eq. (\ref{rre}), this corresponds to relatively small intensities,  $ I_0 \alt 10^{13} Z^{2/3}\lambda_0^{-8/3}$ W/cm$^2$ where the laser wavelength is in $\mu$m. The energy gain rate is shown in Fig. \ref{fig4}. Electrons that are launched in either the parallel or perpendicular directions with respect to the oscillatory electric field lead to similar results. In both cases we observe an enhancement of the energy gain at small velocities.

In order to interpret these numerical results, we first recall the electron scattering theory in the Coulomb field. The particle orbit is uniquely defined by the initial drift velocity, $v_0$, and the impact parameter, $\rho$. The shortest distance along the Kepler trajectory that separates the electron from the ion is $r_{min}=\sqrt{\rho_0^2 + \rho^2} - \rho_0$. In estimates which follow, we assume that the electron can gain a large amount of energy, on the order of $mv_E^2/2$, if it approaches the ion at least as close as $\rho_E$. Then, by assuming $r_{min}=\rho_E$, we find the characteristic impact parameter for these events $\rho_{int}=\sqrt{\rho_E^2+ 2 \rho_E \rho_0}\approx \sqrt{2\rho_E \rho_0}$ for $v_E > v_0$.

During the motion inside the interaction sphere, the electron oscillates many times with a relatively small amplitude, $r_E$, and each time it undergoes a small angle deflection from the original hyperbolic orbit. In an analogy with the Coulomb focusing process \cite{corkum0}, each such deflection brings the electron closer to the ion  and eventually the electron undergoes a large angle scattering which terminates the interaction. As this trajectory modification involves many oscillation periods, it depends weakly on the initial phase. One could estimate the energy gain rate for the case $v_E > v_0$ by taking the product of the number of scattering events per unit time $n_i v_0 \pi \rho_{int}^2$ and the gained energy $m v_E^2$.  That gives the following estimate:
\begin{equation}\label{engw}
    d\epsilon/dt \approx 2\pi n_i Z^2 e^4/m v_0\,.
\end{equation} 
The $1/v_0$ dependence results in large values of the energy gain at small velocities in contrast to the previous theoretical results which do not depend on $v_0$: $d\epsilon/dt \approx 4\pi Z^2 n_i e^4 \Lambda/m  v_E$ \cite{do}. The correlation effects dominate the energy gain in the limit of small electron  velocities, $v_0 <v_E/\Lambda$.

The impact of this mechanism in the enhancement of the electron energy gain is greatly reduced for small intensities where $r_E/\rho_E \ll 1$. Results of Fig. \ref{fig4} have been obtained for the marginal case of $r_E \approx \rho_E$. Still this gain enhancement is quite important as it is independent of the field polarization and the initial phase. That makes the weak field limit distinct from the regime $r_E/\rho_E \gg 1$ which is discussed in two next subsections.

\subsection{Strong field, perpendicular launch}\label{sec4pp}
\begin{figure}[!ht]
\epsfig{figure=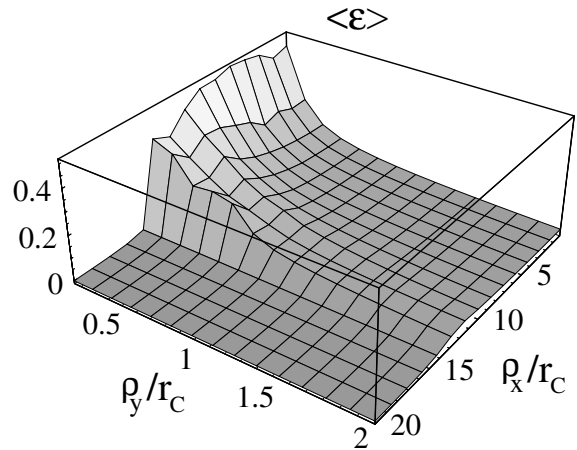, width=7.5cm}
\caption{The average electron energy gain, $\langle\epsilon\rangle$ normalized to the Coulomb energy at the distance $r_C$, $Ze^2/r_C=(Ze^3 E)^{1/2}$, as a function of impact parameters for the perpendicular launch geometry. The energy gain is averaged with respect to the field phases for $v_0 = 0.3 v_E$. The parameters are: $Z=10$, the laser wavelength 0.25 $\mu$m and the intensity $4.43 \times 10^{16}$ W/cm$^2$. The electric field points along the $x$-axis.} \label{fig51}
\end{figure}

The interpretation of high intensity results ($r_E/\rho_E \gg 1$) is more complicated because of the strong dependence on the field phase. We first consider the energy gain of electrons launched transversely to the electric field direction, Fig. \ref{fig5}. Examples of relevant trajectories are presented in Fig. \ref{fig3}c. The energy gain is characterized by two impact parameters: $\rho_x$ --  in the direction the oscillating field and $\rho_y$ -- in the direction of the initial drift velocity. As one can see in Fig. \ref{fig51}, the energy gain is an asymmetric function: it exists over a wide range of impact parameters $\rho_x \alt r_E$, while it is limited to much narrower range $\rho_y \alt \rho_{int}$  in the perpendicular direction.

\begin{figure}[!ht]
\epsfig{figure=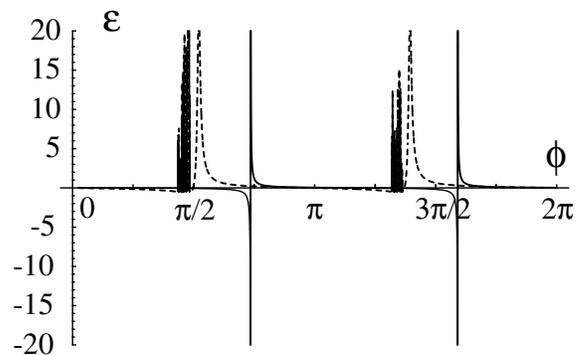, width=7.5cm}
\caption{The electron energy gain, $\epsilon$  normalized to $(Ze^3 E)^{1/2}$ as a function of the initial phase, for the impact parameters $\rho_x = 2.5 r_C$ and $\rho_y = 0$. The perpendicular launch with $v_0 = 0.3 v_E$ (dashed lines) and $3 v_E$ (solid lines). Other parameters are: $Z=10$, the laser wavelength 0.25 $\mu$m and the intensity $4.43 \times 10^{16}$ W/cm$^2$.} \label{fig52}
\end{figure}

Modifications of electron orbits that lead to an enhancement of energy gain also depends on the field phase $\phi$. This is shown in Fig. \ref{fig52}, which presents the dependence of the energy exchange on $\phi$ for two initial drift velocities: the high velocity, $v_0=3 v_E$ (solid line) and the slow initial motion, $v_0=0.3 v_E$ (dashed line). Assuming that incident electrons have a uniform phase distribution, one concludes that only a small fraction of electrons $\approx \rho_E/(v_0/\omega)$ interact strongly with the ion while passing it within the distance of $\rho_E$. A change in the electron energy can be negative or positive. This depends on the direction of electron motion, either towards or away from ion when it enters the sphere of radius $\rho_E$. For fast electrons, the energy loss for particular phases turns out to be exactly the same as the gain for other phases with no net change to the Born approximation results. For slow particles (dashed line in Fig. \ref{fig52}) the reduction of electron energy can result in quasi-trapped trajectories. This motion is irregular and produces a net energy gain when the electron finally escapes the ion. This effect is more important for slower electrons as it leads to increasing energy gain at small drift velocities as shown in Fig. \ref{fig5}. 

An estimate of the energy gain rate $d \epsilon/dt$ involves the characteristic cross section, $\sigma = \pi \rho_x \rho_y$, and the characteristic energy gain $mv_E^2$. The impact parameter, $\rho_x$, along the field direction can be found by assuming that the upper bound for the closest approach and large angle scattering is $r_E$, that is, $r_E=\sqrt{\rho_0^2 + \rho_x^2} - \rho_0$ leading to the estimate $\rho_x = r_E \sqrt{1+2 \rho_0/r_E}$. As discussed above, one have also to account for the fact, that only a small fraction of electrons, $\sim \rho_E/(v_{int}/\omega)$, undergo collisions with this impact parameter. Here, the effective drift velocity $v_{int}=v_0 \sqrt{1+2 \rho_0/r_E}$ is calculated at the distance $r_E$ from the ion. By including these results in the expression for the energy gain, one obtains
\begin{eqnarray} \label{hperp}
\frac{d \epsilon}{d t} &\approx& \pi n_i v_0 (2\rho_0 \rho_E)^{1/2} r_E (1+ 2 \rho_0/r_E)^{1/2} \times \nonumber \\&& \frac{\rho_E \omega}{v_0 (1+2 \rho_0/r_E)^{1/2}}
 \approx \frac{\pi n_i Z^2 e^4}{m v_0}\,.
\end{eqnarray}
This estimate is very similar to Eq. (\ref{engw}) for the energy gain in the weak field limit. The $1/v_0$ dependence explains the growth of the energy gain at small velocities. These are the large angle collisions of selected electrons that contribute to this enhancement. This is moderated by quantum diffraction effects, which are accounted for in the expression for the effective potential (solid line in Fig. \ref{fig5}) if the interaction radius $\rho_{int}$ becomes smaller than the de Broglie length.

\subsection{Strong field, parallel launch}\label{sec4par}
The perpendicular launch considered above represents a general case -- electrons entering the interaction sphere at arbitrary angles with respect to the polarization vector of the oscillatory field are gaining on average the same amount of energy as those launched in the perpendicular direction. Only the electrons propagating parallel to the polarization vector make an exception. As shown in Fig. \ref{fig6}, these electrons are gaining much larger energies at small velocities as compared to the previous cases. The explanation of this anomalous energy gain is related to Coulomb focusing \cite{corkum0} or  the "parachute effect" \cite{fraiman,fraimannew}. Electrons launched in the parallel direction can return to the same ion several times. Each time, as the electron passes the ion, its trajectory is deflected towards the ion. The effective impact parameter decreases until a head-on collision terminates the interaction. The number of possible returns scales as $v_E/v_0$. 

\begin{figure}[!ht]
\epsfig{figure=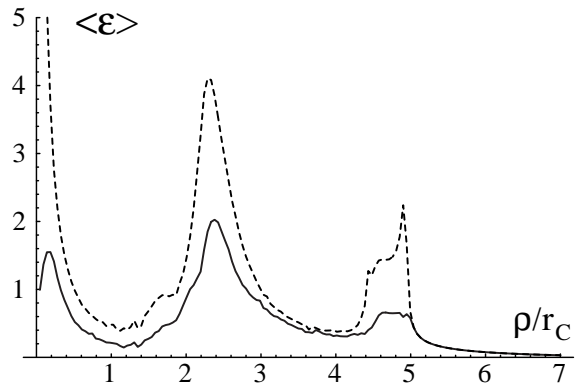, width=7.5cm}
\caption{The average electron energy gain, $\langle\epsilon\rangle$, normalized to $(Ze^3 E)^{1/2}$ as a function of the impact parameter for the parallel launch. Numerical results are averaged with respect to laser field phase for $v_0 = 0.3 v_E$ (dashed line). Other parameters are: $Z=10$, the laser wavelength 0.25 $\mu$m and the intensity $4.43 \times 10^{16}$ W/cm$^2$. Solid lines are obtained from simulations with the effective potential.} \label{fig61}
\end{figure}

\begin{figure}[!ht]
\epsfig{figure=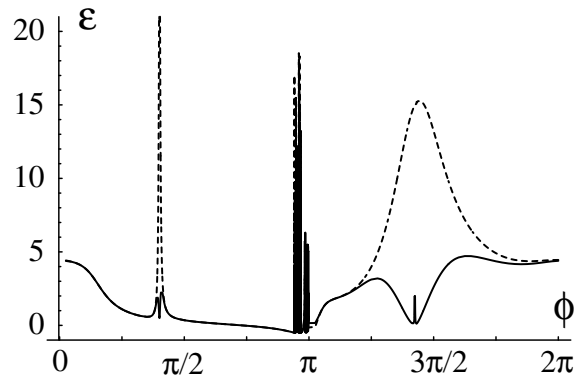, width=7.5cm}
\caption{Dependence of the electron energy gain, $\epsilon$, normalized to $(Ze^3 E)^{1/2}$ on the initial phase for the impact parameter $2.5 r_C$ for the parallel launch and for $v_0= 0.3 v_E$. Other parameters are: $Z=10$, the laser wavelength 0.25 $\mu$m and the intensity $4.43 \times 10^{16}$ W/cm$^2$. Dotted lines correspond to simulations with the effective potential.} \label{fig62}
\end{figure}

This scattering geometry has been analyzed and described in detail by Fraiman et al. \cite{fraiman,fraimannew}. A typical dependence of the energy gain averaged over the field phase $\langle \epsilon\rangle$ (\ref{eq2}) on the impact parameter for the parallel launch is shown in Fig. \ref{fig61} for $v_0=0.3 v_E$. It displays three maxima in the energy exchange corresponding to three electron returns. The second curve (solid line) in Fig. \ref{fig61} has been obtained with the effective potential (\ref{poten}). The amount of energy gained by electrons depends on the short range behavior of the interaction potential at $r \sim \rho_0$ and is reduced when close distance interactions are cut-off (in the present example $\lambda_B > \rho_0$). Figure \ref{fig62} illustrates the sensitivity of the electron energy gain to the field phase and the relative importance of head-on collisions. The latter are represented by the large maximum near $\phi = 3 \pi /2$. The effect of quasi-capture and random trajectories do not strongly contribute to the energy gain in this case. 

An estimate of the heating rate can be made by following the arguments of Fraiman et al. \cite{fraiman,fraimannew}. When passing by an ion, an electron with an impact parameter $\rho$ is scattered by a small angle $\delta \theta \approx 2 \rho_E/\rho$ in the Coulomb potential. This deflection changes the original orbit and the electron acquires a transverse velocity component  $v_\perp= \delta \theta \,v_E$. That electron will arrive into the head-on collision with the ion after half a field period if $\rho = \pi v_\perp/\omega$. From the above expression one finds $\rho = \sqrt{2 \rho_E r_E} \equiv \sqrt{2} r_C$. The electrons with larger impact parameters will also be attracted to the ion but only after several oscillations. Since the number of possible returns is $v_E/v_0$, the characteristic impact parameter for this strong interaction is estimated as $r_C v_E/v_0$. Then the average electron energy gain for the parallel launch can be evaluated as
\begin{equation}\label{epar}
    \frac{d \epsilon}{dt} \approx 2 \pi n_i v_0 \frac{\rho_E}{r_C} r_C^2 \frac{v_E^2}{v_0^2} m v_E^2 = 2 \pi n_i \frac{Z^2 e^4}{m v_0} \left (\frac{r_E}{\rho_E} \right)^{1/2}\,,
\end{equation} 
where we have also accounted for the fact that only fraction of electrons proportional to $\rho_E/r_C$ will undergo strong interactions (cf. Ref. \cite{fraiman}). Although the formula (\ref{epar}) demonstrates the same dependence on the electron drift velocity as (\ref{hperp}), the energy gain is $(r_E/\rho_E)^{1/2} \gg 1$ times bigger due to the parachute effect discussed above. This Coulomb attraction due to multiple collisions is effective, however, only for small angles between ${\bf v}_0$ and the electric field vector, $\theta \alt \sqrt{\rho_E/r_E}$, where the velocity $v_\perp$ has a well defined sign. For larger angles, the electron-ion interaction reduces to a single collision event, as presented previously for the case of perpendicular launch.

\section{Heating rate for an isotropic electron distribution}\label{sec43}
The energy gain rates (\ref{hperp}) and (\ref{epar}) calculated for mono-energetic electrons can be used for evaluating heating rates for various electron distribution functions. Here we consider the common case of an isotropic Maxwellian electron distribution function which is established  due to electron-electron and electron-ion collisions. Before analyzing the statistical significance of correlated collisions, we recall the basic elements of the classical theories.

By taking an average with respect to a Maxwellian distribution function with the temperature $T_e$, the Dawson-Oberman theory (\ref{hrborn}) gives the following formula for the heating rate per electron:
\begin{equation}\label{eqMaxclass}
\frac {d T_e}{d t} =  \sqrt{2 \pi} \,\frac{8n_i Z^2 e^4 v_E^2}{m v_{Te}^3} \sum_{l=1}^{+\infty} l^2 \int_0^{q_m} \frac {d q}{q^4} I_l(q) \exp \left( -\frac{l^2 mv_E^2}{2 q^2 T_e}\right )\,,
\end{equation}
where  $I_l(x)=\int_0^x J_l^2(y)dy$, $v_{Te}=\sqrt{T_e/m}$ is the electron thermal velocity, and the parameter $q_m=2\hbar \omega/T_e$ defines the cut-off at small distances. The quantum mechanical theory  (\ref{hrkw}) gives the following heating rate 
\begin{eqnarray}\label{eqMaxquan}
\frac {d T_e}{d t} &=&\sqrt{2 \pi} \frac{16 n_i Z^2 e^4 v_E^2}{\hbar \omega v_{Te}} \sum_{l=1}^{+\infty} l \sinh{\frac{l \hbar \omega}{2 T_e}} \int_0^\infty \frac {d q}{q^4} I_l(q) \times  \nonumber \\ &&
\exp{\left(-\frac{l^2 mv_E^2}{2 q^2 T_e}-\frac{q^2 \hbar^2 \omega^2}{8m v_E^2 T_e}\right)},
\end{eqnarray}
One can define the effective collision frequency by the relation: $ d T_e/dt = \nu_{eff} m v_E^2/2$ and use Eqs. (\ref{eqMaxclass}) and (\ref{eqMaxquan})  to evaluate it. It is well-known that in the weak field limit, $v_E \ll v_{Te}$, the effective collision frequency coincides with the electron-ion transport  collision frequency $\nu_{ei}=4 \sqrt{2 \pi} Z^2 e^4 n_i \Lambda/3 m^2 v_{Te}^3$, where the Coulomb logarithm $\Lambda$ is defined by Eq. (\ref{eqLambda}). In the strong field limit, $v_E \gg v_{Te}$, $\nu_{eff}$ decreases as $v_E^{-3}$ and there is a simple expression from Ref. \cite{vick,faehl} which provides a very good interpolation for all existing analytical theories:
\begin{equation}\label{vick}
\nu_{eff}=\nu_{ei} (1+ v_E^2/6 v_{Te}^2)^{-3/2}\,.
\end{equation}

\begin{figure}[!ht]
\epsfig{figure=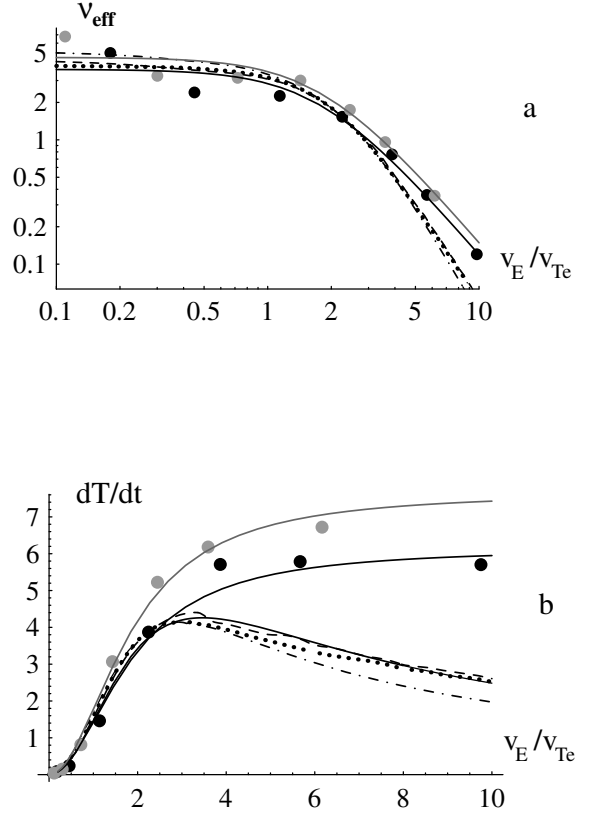, width=7.5cm}
\caption{Dependence of the effective collision frequency normalized to the electron-ion collision frequency $\nu_{ei}=4 \sqrt{2 \pi} n_i Z^2  e^4/3 m^2 v_{Te}^3$ (a) and the absorption rate per electron $dT_e/dt$ normalized to $\nu_{ei} T_e$ (b) on $v_E/v_{Te}$ for a plasma with $Z=10$ and for the Maxwellian electron distribution function with $T_e=200$ eV (black dots) and 500 eV (grey dots). The continuous lines following large dots are calculated from the formula (\protect \ref{hMax}). Numerical results are compared with the theoretical results for $T_e=200$ eV: the Kroll-Watson approximation -- dash-dotted line, the Dawson-Oberman approximation -- dashed line, the classical approach -- dotted line, and the expression (\protect \ref{vick}) -- the fine solid line. The laser wavelength is 0.25 $\mu$m.} \label{fig7}
\end{figure}

Figure \ref{fig7}a compares the variation of effective collision frequency with $v_E/v_{Te}$ found from our simulations and from theoretical formulas presented above. The numerical heating rates were calculated by  the direct average of  the energy gain rate over the isotropic Maxwellian distribution function. Numerical results have been obtained for the cases when the electron temperature is 200 eV and also for 500 eV. All models and numerical simulations generally agree for $v_E \alt v_{Te}$. Some discrepancies at small values of $v_E$ follow from differences between theoretical approximations, in particular, from the treatment of short range interactions. 

The most dramatic difference between the known theories and our results is illustrated in Fig. \ref{fig7}b which shows that the calculated heating rate reaches a constant values in the large field limit, $v_E > v_{Te}$. Our results for the effective collision frequency can be approximated by the following formula:
\begin{equation} \label{hMax}
\nu_{eff}=\nu_{ei} (1+ 0.3 \, v_E^2/v_{Te}^2)^{-1}\,.
\end{equation}
This expression gives a $1/v_{Te}$ dependence in the heating rate for the limit of large values of $v_E/v_{Te}$. This agrees with the conclusions of Ref. \cite{fraiman}. Such behavior agrees with the $1/v_0$ dependence of the energy gain rate for the perpendicular electron launch discussed in  sections \ref{sec4w} and \ref{sec4pp}. The contribution of electrons with parallel velocities is not evident in Eq. (\ref{hMax}) because small angles, $\theta \alt \sqrt{\rho_E/r_E}$, make a small statistical contribution to the average quantity for the isotropic distribution function. 

We recall that the enhancement of the heating rate due to the correlation effect takes place in a sufficiently rarified plasma where the electron quiver radius $r_E$ is smaller than the inter-particle distance $d$. The condition (\ref{vcon1}) ensures this limitation.

\section{Heating rates for an anisotropic electron distribution}\label{sec42}
It was shown in Sec. \ref{sec4par} that slow electrons ($v_0<v_E$) propagating along the oscillatory field are the main contributors to the enhanced heating rates. However this effect is confined to small angles and it disappears if electrons are distributed isotropically. Therefore for typical laser plasma interaction conditions, the heating rate might depend dramatically on the number of slow electrons and on the anisotropy of actual electron distribution function. In particular, this effect could be important for plasmas produced by the ionization of gases by high intensity laser pulses. According to Refs. \cite{corkum1,bychenk}, the velocity distribution function of the photoelectrons displays a strong anisotropy along the direction of the linearly polarized laser beam, if tunnel ionization is the dominant mechanism. The following simple analytical expression captures the main features of electrons produced by the tunnel ionization process \cite{bychenk}:
\begin{equation}\label{adist}
F_0({\bf v}) = n_e (m /2 \pi T_e)^{1/2}\delta({\bf v}_\perp)\exp(-m v_\|^2/2 T_e)
\end{equation}
where $v_\|$ is the electron velocity in the direction of laser polarization and the effective electron temperature $T_e$ may be a function of the ionization potential and the laser field amplitude \cite{bychenk}. In our analysis we treat $T_e$ as a free parameter.

We compare results of various analytical models with our simulations after averaging them with respect to the distribution function (\ref{adist}). This averaging reduces to a simple integration of expressions for the electron energy gain in the parallel geometry (\ref{bornpar}) and (\ref{kwpar})  with a one-dimensional Maxwellian velocity distribution. The main contribution to this integral comes from slow electrons. In the case where $v_E< v_{Te}$, all classical theories give similar results that depend on the cut-off parameters. For the quantum approach, the heating rate per electron derived from Eq. (\ref{kwpar}) has the following form:
\begin{eqnarray}
\frac {d T_e} {d t} &= &n_i Z^2 e^4 \sqrt{\frac {\pi m}{2 T_e}} \sum_{l=-l_{min}}^{+\infty}\ l \xi_{E} \int \limits_{-\infty}^{+\infty} du \exp \left(-\frac{mv_E^2}{2T_e} u^2 \right) \times \nonumber \\ &&
\int \limits_{u-\sqrt{u^2+\xi_{E} l}}^{u+\sqrt{u^2+\xi_{E} l}} d y \frac{ J_l^2(2y/\xi_{E})}{(y u+ \xi_{E} l/2)^2},
\end{eqnarray}
where $\xi_{E}=2\hbar \omega/m v_E^2$.

\begin{figure}[!ht]
\epsfig{figure=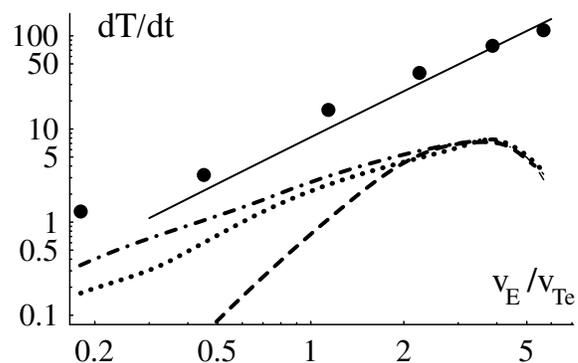, width=7.5cm}
\caption{Dependence of the absorption rate per electron $dT_e/dt$ normalized to $\nu_{ei} T_e$ on the electron quiver velocity for a laser wavelength 0.25 $\mu$m and for an anisotropic distribution function (\protect\ref{adist}) with $T_e=200$ eV: numerical results -- dots, the Kroll-Watson approximation -- dash-dotted line, the Dawson-Oberman approach -- dashed line, the classical approach -- dotted line, expression (\protect\ref{nuan}) -- solid line.} \label{fig8}
\end{figure}

Figure \ref{fig8} shows the intensity dependence of the heating rates that are obtained from our numerical model and from various analytical expressions. The striking discrepancy between the analytical results and our numerical simulations is a consequence of the large number of slow electrons in the electron distribution function (\ref{adist}) propagating parallel to the  polarization direction. Contrary to the case of a three dimensional Maxwellian distribution function, expression (\ref{adist}) contains a finite number of electrons at zero velocity and therefore the heating rate diverges as a logarithm of $v_{min}$, if one averages the energy gain (\ref{epar}) with the one-dimensional electron distribution function. This divergence can be resolved by accounting for the fact that the maximum impact parameter $\rho_{max} \approx r_C\, v_E/v_0$ cannot be greater than the inter-particle distance $d$. This leads to the following estimate $v_{min} \approx v_E r_C/d$. Then one finds the following expression for the heating rate per electron:
\begin{equation}\label{nuan}
    \frac{dT_e}{dt} =\frac{\sqrt{2 \pi} n_i Z^2 e^4}{m v_{Te}}\, \left( \frac {r_E}{\rho_E} \right)^{1/2}\, \ln \left(d \sqrt{\frac{T_e}{Ze^2r_E}}\right)\,.
\end{equation} 
This expression is proportional to $v_E^{3/2}$. It approximates well numerical results as shown with the solid line in Fig. \ref{fig8}. 

Photo-ionized plasmas with anisotropic electron distributions provide an example of a case where enhanced heating rates are obtained due to correlated collisions as described in our paper. It is clear that the particular form (\ref{adist}) of the electron distribution function can be quickly altered due to rapid heating. However, inverse bremsstrahlung heating can potentially support a certain level of anisotropy in the electron distribution function. More work needs to be done in developing the self-consistent electron distribution for this case and in comparing this model with experiments.

\section{Conclusions}\label{sec5}
We have performed numerical simulations based on the test particle model which involves classical electron trajectories in the electric field of an ion and a homogeneous, high frequency laser field. We have confirmed recent results by Fraiman {\it et al}. \cite{fraiman} regarding the electron orbits that closely approach an ion several times during each scattering event and subsequently lead to a large angle deflection and to a large enhancement in the electron energy. These correlated collisions have been ignored in preceding theories which significantly underestimate the electron heating rates by neglecting large angle scattering and the important field phase dependence of electron trajectories. The inverse-bremsstrahlung heating rates have been evaluated for different electron distribution functions, including: mono-energetic beam-like electrons, Maxwellian and highly anisotropic distribution relevant to photo-ionized gases. The energy gain for electrons with small drift velocities and for large amplitude electric fields, $v_0 < v_E$, increases as $1/v_0$ and exceeds the previously known results if $v_0 < v_E/\Lambda$. This behavior contradicts results of all analytical theories and it is especially important for electrons with initial velocities that are parallel to the field amplitude. When the quantum short-distance cut-off is introduced through an effective potential, the energy gains demonstrate a similar dependence on the particle drift velocity but a smaller magnitude of enhancement is obtained if the cut-off length, $\lambda_B =2\pi \hbar/mv_E$, is larger than the characteristic radius of the Coulomb interaction $\rho_C=Ze^2/T_e$.

For isotropic Maxwellian distribution functions, the heating rate in the limit of $v_E > v_{Te}$ is \textit{independent} on the laser intensity. It can be approximated by the interpolation expression (\ref{hMax}) for the effective collision frequency. For highly anisotropic electron distribution functions such as those produced by photo-ionization, the new effect of large angle correlated scattering dominates the heating rate at high laser intensities. It leads to $v_E^{3/2}$ dependence of the heating rate.  

Our numerical results involve calculations of the electron scattering on a single ion. They are relevant to realistic plasma conditions provided the two-body collision approximation $r_E<d$ (\ref{vcon1}) is satisfied. This imposes an upper limit on the laser intensity for a given ion density and typically holds for very underdense plasmas. Also, we have implicitly assumed that the plasma is weakly coupled and that the average distance between ions, $d$ is smaller than the screening length, $\lambda_D$.

The enhancement of the inverse bremsstrahlung heating rates should influence  laser plasma interaction experiments. For example, the collisional heating of long preformed plasmas by high intensity short laser pulses accompanies several important applications, including X-ray lasers, laser wake-field particle accelerators, and the high-harmonics generation. Moderate ion densities in the plasma waveguide ($n_i=10^{17} - 10^{19}$ cm$^{-3}$) should make it easier to satisfy condition (\ref{vcon1}). High intensity picosecond laser pulses would allow investigation of the heating rate dependence on $v_E/v_{Te}$ over the wide range of values. Atomic processes involved in X-ray lasers are very sensitive to plasma temperature and could provide a diagnostic method for the inverse bremsstrahlung heating rates in addition to other diagnostics techniques such as the Thomson scattering and absorption measurements. 

The authors thank Dmitri Romanov for his advise concerning numerical calculations. This work was supported by the Natural Sciences and Engineering Research Council of Canada, the Alberta Ingenuity Fund and the Russian Foundation for Basic Research (grant N 00-02-16063).


\begin{thebibliography}{99}
\bibitem{do} J. Dawson, C. Oberman, Phys. Fluids {\bf 5}, 517 (1962).
\bibitem{silin} V. P. Silin, Sov. Phys. JETP {\bf 20}, 1510 (1965).
\bibitem{bunkin}  F. B. Bunkin, A. E. Kazakov, and M. V. Fedorov,  Sov. Phys. Uspekhi {\bf 15}, 416 (1972); F. V. Bunkin and M. V. Fedorov, Sov. Phys. JETP {\bf 22}, 844  (1966). 
\bibitem{kw} N. M. Kroll and K. M. Watson, Phys. Rev A {\bf 8} , 804 (1973).
\bibitem{kull} H.-J. Kull and L. Plagne, Phys. Plasmas {\bf 8}, 5244 (2001).
\bibitem{pert1} G. J. Pert, J. Phys. A: Gen. Phys. {\bf 5}, 506 (1972).
\bibitem{pert2} G. J. Pert, J. Phys. B: Atom. Molec. Phys. {\bf 12}, 2755 (1979). 
\bibitem{decker} C. D. Decker, W. B. Mori, J. M. Dawson, and T. Katsouleas, Phys. Plasmas {\bf 1}, 4043 (1994) 
\bibitem{shvets} G. Shvets, N. J. Fish, Phys. Plasmas {\bf 4}, 428 (1997).
\bibitem{mulser} P. Mulser, F. Cornolti, F. B\'esuelle, and R. Schneider, Phys. Rev. E {\bf 63}, 016406 (2001). 
\bibitem{corkum0} P. B. Corkum, Phys. Rev. Lett. {\bf 71}, 1994 (1993).
\bibitem{brabec} T. Brabec, M. Yu. Ivanov and P. Corkum, Phys. Rev. A {\bf 54}, R2551 (1996).
\bibitem{ivanov} G. L. Yudin and M. Yu. Ivanov, Phys. Rev. A {\bf 63}, 0330404 (2001).
\bibitem{fraiman} G. M. Fraiman, V. A. Mironov, and A. A. Balakin, Phys. Rev. Lett. {\bf 82}, 319 (1999); JETP {\bf 88}, 254 (1999);
G. M. Fraiman, A. A. Balakin, and V. A. Mironov, Phys. Plasmas {\bf 8}, 2502 (2001).
\bibitem{fraimannew}A. A. Balakin and G. M. Fraiman, JETP {\bf 93}, 695 (2001). 
\bibitem{fedorov} M. V. Fedorov, {\it Atomic and Free Electrons in a Strong Light Field}, World Scientific (Singapore, 1997). 
\bibitem{ichimaru} S. Ichimaru, {\it Statistical Plasma Physics}, Addison-Wesley (Redwood City, 1992).
\bibitem{num1} L. Wiesenfeld, Phys. Lett. A {\bf 144}, 467 (1990).
\bibitem{jetp} N. L. Manakov, A. F. Starace, A. V. Flegel, and M. V. Frolov, JETP Lett. {\bf 76}, 258 (2002).
\bibitem{dufty} B. Talin, A. Calisti, J. Dufty, Phys. Rev. E {\bf 65}, 056406 (2002).
\bibitem{batishchev} O. V. Batishchev, A. Brantov, V. Yu. Bychenkov, W. Rozmus, R. Sydora, and C. E. Capjack, Bull. American Phys. Soc. {\bf 47} (9), 230 (2002).
\bibitem{deutsch} C. Deutsch, Phys. Lett. A {\bf 60}, 317 (1977).
\bibitem{uhlenbeck} G. E. Uhlenbeck, L. Gropper, Phys. Rev. {\bf 41}, 79 (1932).
\bibitem{catto} P. J. Catto and Th. Speziale, Phys. Fluids {\bf 20}, 167 (1977). 
\bibitem{pert3} G. J. Pert, Phys Rev. E {\bf 51}, 4778 (1995).
\bibitem{ferrante} G. Ferrante, C. Leone, and L. LoCasio, J. Phys. B {\bf 12}, 2319 (1979). 
\bibitem{landau3} L. D. Landau and E. M. Lifshitz, {\it Quantum Mechanics (Nonrelativistic Theory)}, 3rd edition, Pergamon Press (Oxford, 1976).
\bibitem{schlessinger} L. Schlessinger and J. Wright,  Phys Rev. A {\bf 20}, 1934 (1979). 
\bibitem{ichimaru1} S. Ichimaru, {\it Statistical Plasma Physics}, vol. II: {\it Condensed Plasmas}, Addison-Wesley (Reading, 1994).
\bibitem{comment} So-called the low frequency or Krall-Watson approximation \cite{kw}, is often refered to as the quantum calculation improving on the Born approximation. However the limits of applicability of this theory were not well defined in the literature. As discussed by Fedorov \cite{fedorov} (pp. 109 - 110), the low frequency approximation works correctly only in the regime that corresponds to the Born approximation. 
\bibitem{candy} J. Candy, W. Rozmus, J. Comp. Phys. {\bf 92}, 230 (1991).
\bibitem{daniel} R. Daniele, F. Trombetta, G. Ferrante, P. Cavaliere, and F. Morales, Phys Rev. A {\bf 36}, 1156 (1987). 
\bibitem{faehl} R. J. Faehl and N. F. Roderick, Phys. Fluids {\bf 21}, 793 (1978). 
\bibitem{vick} D. Vick, C. E. Capjack, V. T. Tikhonchuk, and W. Rozmus, Comments Plasma Phys. Cont. Fusion {\bf 17}, 87 (1996).  
\bibitem{corkum1} P. B. Corkum, N. H. Burnett, and F. Brunel, Phys. Rev. Lett. {\bf 62}, 1259 (1989); N. H. Burnett  and P. B. Corkum, J. Opt. Soc. Am. {\bf B6}, 1195 (1989). 
\bibitem{bychenk} V. Yu. Bychenkov and V. T. Tikhonchuk, Laser Physics {\bf 2}, 525 (1992). 
\end{thebibliography}
\end{document}